\documentclass{PoS}

\title{KYNREFREV - implementation of an X-ray reverberation model in XSPEC}

\ShortTitle{X-ray reverberation}

\author{\speaker{M. D. Caballero-Garcia$^a$}, M. Dov\v{c}iak$^a$, I. Papadakis$^{bc}$, A. Epitropakis$^b$, J. Svoboda$^a$, E. Kara$^d$, V. Karas$^a$\\
\\
 \llap{$^a$} Astronomical Institute of the Academy of Sciences, Bo\v{c}n\'{\i} II 1401, CZ-14100 Praha 4, \\
             Czech Republic, E-mail: \email{garcia@asu.cas.cz}                \\
 \llap{$^b$} Department of Physics and Institute of Theoretical and Computational Physics, \\
             University of Crete, GR-71003 Heraklion, Greece \\
 \llap{$^c$} Foundation for Research and Technology - Hellas, IESL, Voutes, GR-7110 Heraklion, \\
             Greece \\
 \llap{$^d$} Department of Astronomy, University of Maryland, College Park, MD 20742-2421, USA \\

}

\abstract{We present briefly the first results obtained by the application of the KYNREFREV-reverberation model, which is ready for its use in XSPEC. This 
model computes the time dependent reflection spectra of the disc as a response to a flash of primary power-law radiation from a point source corona located on the axis 
of the black-hole accretion disc. The assumptions of the model are: central Kerr black hole, surrounded by a
Keplerian, geometrically thin, optically thick, ionised disc with the possibility of defining the radial density profile and a stationary hot point-like patch of plasma located
on the system rotation axis and emitting isotropic power-law radiation (lamp-post geometry). Full relativistic ray-tracing code in vacuum is used for photon paths from the corona
to the disc and to the observer and from the disc to the observer. The ionisation of the disc is set for each radius according to the amount of the
incident primary flux and the density of the accretion disc. In this paper we comment on some preliminary results obtained through the analysis of X-ray reverberation time-lags
from 1H~0707-495 and IRAS~13224-3809.
}

\FullConference{Accretion Processes in Cosmic Sources - APCS2016 -\\
		5-10 September 2016,\\
		Saint Petersburg, Russia}

\begin{document}

\section{Introduction}

The X-ray emission from accreting black holes (BHs), in particular the narrow line Seyfert 1 (NLS1) Active Galactic Nuclei (AGN) reported in this work, is believed to 
come from the innermost part of the accretion disc. This has been
seen to be highly variable \cite{leighly99,turner99,ponti12}. The accretion disc reprocesses and ``reflects'' the inner radiation, so part of the X-ray photons emitted by the corona
are reflected and vary in luminosity partially according to variations in the flux received from the latter. The emission at the energy bands predominantly composed of reflection
from the accretion disc (i.e. the soft-excess at 0.3-1\,keV and the broadened Fe K$_{\alpha}$ line at ${\approx}$3-8\,keV) is expected
to lag behind correlated variations in energy bands dominated by the directly observed continuum emission (1-4\,keV). The time delay corresponds to the additional light travel time
from the coronal source to the reflecting accretion disc \cite{uttley14}. Some of these sources show a reflection-dominated spectrum {\it and potentially have maximally spinning BHs}. This fact
can be understood if light bending is an important effect for them \cite{miniutti04}. Their high-energy source (i.e. the jet or
corona) is very close to the BH and the observed direct continuum (i.e. power-law) is very low. This would correspond to Regime I
of \cite{miniutti04}, corresponding to a low height of the primary source (${\rm h}{\le}2-4\,{\rm r}_{\rm g}$, depending on the observer inclination), where strong light-bending suffered 
by the primary radiation dramatically reduces the observed power-law emission component at infinity and the total luminosity at infinity stays relatively constant. During Regime II
the height of the primary source is higher (${\rm h}{\approx}4-13\,{\rm r}_{\rm g}$, depending on the observer inclination) and the total luminosity at infinity varies by a factor of four
and mostly due to variations of the primary source.

The recent detection of time delays, as a function of the Fourier frequency, between the soft-band and the hard-band photons in Active Galactic Nuclei (AGN), can 
shed light on the X-ray emission mechanism and the geometry of these BH systems. Currently, the observed negative X-ray time-lags (i.e. soft-band variations lagging the hard band
variations observed at high frequencies) have triggered a great deal of scientific interest on interpreting their nature. The first tentative detection was in the 
AGN Ark 564 \cite{mchardy07}, where an
origin in reflection from the accretion disc was proposed. The first statistically significant detection came from 
the AGN 1H~0707-495 \cite{fabian09}. More recent works \cite{emmanoulopoulos11,demarco13} then found that such X-ray time-delays are 
much more common than it was initially thought by the analysis of the X-ray data from several tens of AGN \cite{kara16}. The opposite time-delayed behaviour,
i.e. positive X-ray time-delays (hard-band variations lag the soft band variations observed at low frequencies), has been known for quite some time in both 
AGN (e.g. \cite{papadakis01,mchardy04,arevalo06})
and X-ray binaries (e.g. \cite{miyamoto89,nowak96,nowak99}). Although positive time-lags are expected in the standard Comptonization process within the X-ray source \cite{nowak99},
they can also be produced by diffusive propagation of perturbations in the accretion flow \cite{kotov01}.

In this paper we briefly describe a generalized scheme and a code which can be used with the standard X-ray spectral fitting package, XSPEC \cite{arnaud96}. This model takes into
account a treatment of all the general relativistic effects for accretion discs and light bending effects in the disc-corona geometry. Time-lags versus frequency and energy can 
be produced and fitted by the user, either in XSPEC or using another suitable tool. Sec.~\ref{mod} shows a brief description of the model, Sec.~\ref{results} provides some
preliminary results obtained using the model, and in Sec.~\ref{discuss} we briefly discuss and compare them with other results obtained in the literature.

\subsection{The model}  \label{mod}

This model ({\tt KYNREFREV}; \cite{dovciak17}) computes the X-ray emission from an accretion disc that is illuminated from the primary power-law source located on the axis above the
central BH with a ``flash'' (hereafter referred to as ``flare''). All relativistic effects are taken into
account for photon paths in vacuum from the corona to the disc and to the observer and from the disc to the observer. The transfer functions (see \cite{dovciak04}) for 
these effects have been precalculated and stored in the FITS files that are used by the model to speed up the execution of the code. The reflection is taken from the 
REFLIONX\footnote{https://heasarc.gsfc.nasa.gov/xanadu/xspec/newmodels.html} FITS table files \cite{rossfabian05}, which are
integrated over the disc.

\begin{figure}
\includegraphics[bb=-85 0 464 135,angle=0,width=1.5\textwidth]{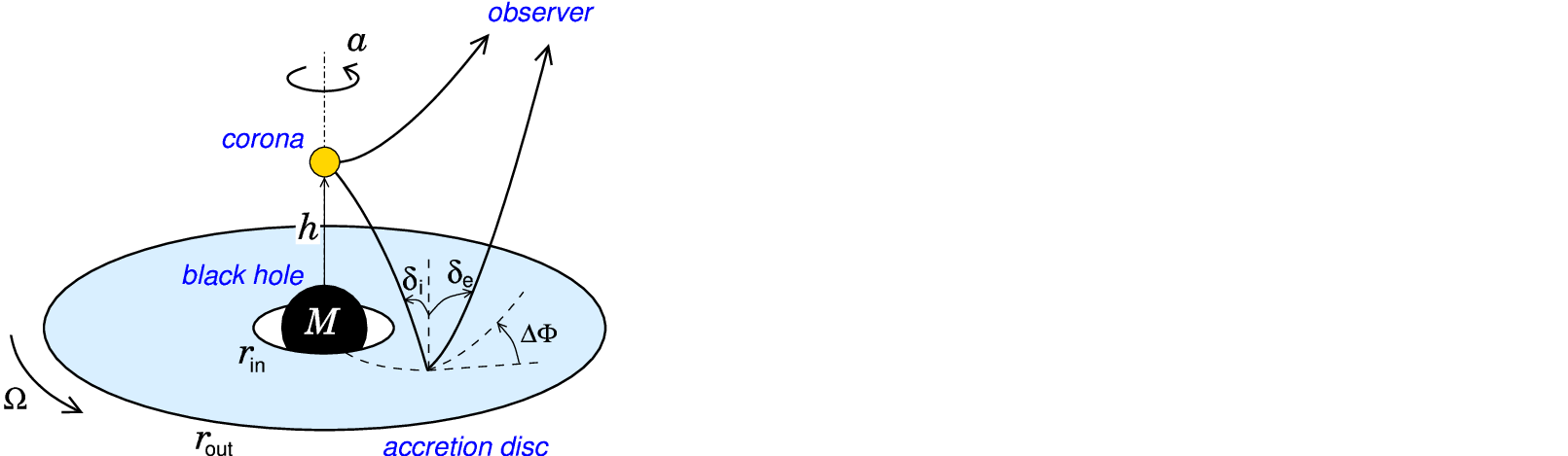}
\caption{The sketch of the lamp-post geometry.}
\label{fig1}
\end{figure}

The lamp-post geometry used in the KYNREFREV model (see Fig.~\ref{fig1}) consists of a BH, surrounded by an equatorial accretion disc, that is illuminated by an X-ray source
located on the symmetry axis of the disc. Some of the main physical parameters of the model are the mass (in units of $10^8{\rm M}_{\odot}$) and the angular momentum ($-1{\le}{\rm a}/{\rm M}{\le}1$), i.e.
the spin of the BH in geometrized units, the height (h) of the X-ray source (in units of ${\rm r}_{\rm g}={\rm GM}/{\rm c}^2$), and the viewing angle (${\theta}$) of a distant observer with
respect to the axis of the disc. 

The disc is assumed to be ionized, geometrically thin, Keplerian and co-rotating with the BH, with a radial extent ranging from the inner edge of the disc (in units of the innermost stable
circular orbit, ${\rm r}_{\rm ISCO}$) up to the outer radius (${\rm r}_{\rm out}=10^{3}\,{\rm r}_{\rm g}$). The spin of the BH uniquely defines ${\rm r}_{\rm ISCO}$. When
measured in geometrized units, the spin can attain any value between zero and unity, with $a=0$ (${\rm r}_{\rm ISCO}=6\,{\rm r}_{\rm g}$) and $a=1$ (${\rm r}_{\rm ISCO}=1\,{\rm r}_{\rm g}$)
indicating a non-spinning (i.e. Schwarzschild) and a maximally spinning (i.e. ``extreme'' Kerr) BH, respectively. The primary X-ray source (i.e. corona) is assumed to be a hot cloudlet of plasma on
the rotation axis at height $h$ and emitting a power-law spectrum of X-ray radiation, ${\rm F}_{\rm p}{\propto}{\rm E}^{-{\Gamma}}\mathrm{e}^{{\rm E}/{\rm E}_{\rm c}}$, with
a sharp low-energy cut-off at 0.1\,keV and high-energy exponential cut-off at ${\rm E}_{\rm c}=300$\,keV.

In the new model the REFLIONX \cite{rossfabian05} tables for constant density slab illuminated by the power-law incident radiation are used to compute the local reprocessing of the incident radiation in the ionized 
accretion disc. The ionization
of the disc, ${\xi}$, is set by the amount of the incident primary flux (dependent on the luminosity and height of the primary source and the mass of the BH) and by the density of
the accretion disc as ${\xi(r)}{\propto}{\rm F(r)}/{\rm n(r)}$. We used a constant-density disc in our analysis. This simplification is well satisfied because the radial dependence of any realistic density
profile is much less significant than the radial decrease of the disc illumination by the lamp-post corona (see, e.g., Fig.~3 in \cite{svoboda12}).

We may consider several limb brightening/darkening prescriptions for the directionality of the re-processed emission: isotropic
emission (flux${\sim}$1), Laor's limb darkening (flux${\sim}(1+2.06{\mu}$); \cite{laor91}) and Haardt's limb brightening (flux${\sim}ln(1+1/{\mu})$; \cite{haardt93,czerny04}), with ${\mu}$ being cosine of 
emission angle.

\subsubsection{Transfer function and time-lags}

We refer as reverberation to the variable signal that we get (i.e. hereafter called ``response function'') corresponding
to the variations of the driving primary source of radiation \cite{uttley14}. We assume that
the emission which we see in each band is related to the underlying driving signal by a linear impulse response, which represents
the response of the emission in that band to an instantaneous ``flash'' (i.e. a delta-function impulse) in
the underlying primary signal. Therefore, the observed light curve is the convolution of the underlying driving signal time-series with the impulse response for that energy band.

In this work we refer as ``transfer function'' the {\it relative response} of the disc to the illumination:

\[
{\psi}_{\rm r}(E,t)=\frac{{\rm F}_{\rm r}}{{\rm F}_{\rm p}} ,
\]

\noindent where ${\rm F}_{\rm r}(E,t)$ is the time dependent observed reflected flux from the disc as a response to a flare\footnote{Implicitly referred to a ``Box'' flare in practice because of numerical/computation reasons.} that
would be observed as ${\rm F}_{\rm p}{\delta}(t)$ (${\rm F}_{\rm p}$ is total observed primary flux at energy E).

\noindent The Fourier transform of the transfer function is calculated as:

\[
\hat{{\phi}}_{\rm r}(E,f)={\rm A}_{\rm r}(E,f)\mathrm{e}^{{\phi}_{\rm r}(E,f)}
\]

\noindent with amplitude ${\rm A}_{\rm r}(E,f)$ and phase ${\phi}_{\rm r}(E,f)$ (which is sometimes referred to as transfer function in other works).

And the total phase of the Fourier transform of the total flux (i.e. both the primary and the reflection continuum are taken into account) is:

\[
{\phi}_{\rm tot}(E,f)=\arctan{\frac{{\rm A}_{\rm r}(E,f)\sin{{\phi}_{\rm r}(E,f)}}{1+{\rm A}_{\rm r}(E,f)\cos{{\phi}_{\rm r}(E,f)}}} .
\]

\noindent One can calculate the time-lag of the signal, computed from the total phase at energy bin E with respect to the total phase at some reference energy bin:

\[
{\tau}(E,f)=\frac{{\Delta}{\phi}_{\rm tot}(E,f)}{2{\pi}f} ,
\]

\noindent where ${\Delta}{\phi}_{\rm tot}(E,f)$ is defined as the phase difference at any energy value $E$ with respect to the reference energy range $E_{\rm ref}$, i.e.
${\Delta}{\phi}_{\rm tot}(E,f)={\phi}(E,f)-{\phi}(E_{\rm ref},f)$.

To determine the response function of the disc, we assume that the primary X-ray source isotropically emits a flare of duration
equal to 1\,${\rm t}_{\rm g}$ (${\rm t}_{\rm g}={\rm GM}/{\rm c}^3$). However, after computing the response, the Fourier transform is corrected for the ``box'' shape of the primary flare 
so that it corresponds to the delta function impulse. Upon being illuminated, each area element of the disc ``responds'' to this flare by isotropically and instantaneously
emitting a ``reflection spectrum'' in its rest-frame. We assume that the reprocessed flux is proportional to the incident
flux, and that the disc material is ionized.

\section{Analysis and results} \label{results}

\subsection{1H~0707-495}

1H~0707-495 (z=0.0411) is a narrow-line Seyfert 1 galaxy with a BH mass estimate of ${\rm M}=2{\times}10^6\,{\rm M}_{\odot}$ \cite{bianandzhao03}. It is one of the most highly variable Seyfert galaxies, commonly
known to increase by an order of magnitude in flux on timescales as short as one hour. Therefore this makes this source an ideal target for extensive spectral-timing analysis as previous studies based on the X-ray
data from the {\it XMM-Newton} satellite have shown so far. \cite{fabian09} and \cite{zoghbi10} showed that the high-frequency variations in the soft excess from 0.3-1\,keV lag behind those in the 1-4\,keV band 
by $\sim 30$\,s. They interpreted this short 
timescale lag as the reverberation time delay between the primary-emitting corona and the inner accretion disc. This short time delay would put the corona at a height of $<10\,{\rm r}_{\rm g}$ from the accretion disc
(using the aforementioned value for the BH mass).

\subsection{IRAS~13224-3809}

IRAS~13224-3809 (z=0.066) is also one of the most X-ray variable Seyfert 1 galaxies known \cite{boller96}, and therefore it is a useful source for which to probe the environments of the innermost 
regions around the BH. It was first observed with {\it XMM-Newton} in 2002 for 64\,ks \cite{boller03,gallo04,ponti10}, and later for 500~ks, which led to a significant detection of the soft 
time-lag \cite{fabian12}. Later \cite{kara13} followed the source and reported on the analysis of the soft time-lags of IRAS~13224-3809 in the low and high flux states. Together with 1H~0707-495 it is one of the very 
few sources for which {\it XMM-Newton} has devoted long-exposure (${\gtrsim}$500\,ks) observational campaigns.

\subsection{Time-lag versus frequency spectral fitting}

\begin{figure}
\includegraphics[bb=0 -200 612 292,angle=270,width=0.4\textwidth]{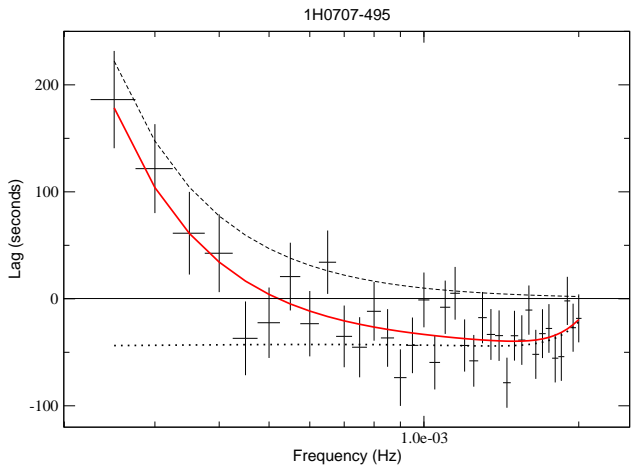}
\caption{The (0.3-1 versus 2-4\,keV) X-ray soft time-lag versus frequency spectrum of 1H~0707-495 fitted with the {\tt KYNREVREF} model in XSPEC (red line). The two components of the best-fitting time-lag model are the 
relativistic reflected component (black-dotted line) and the power-law (black-dashed line).}
\label{fig2}
\end{figure}

\begin{figure}
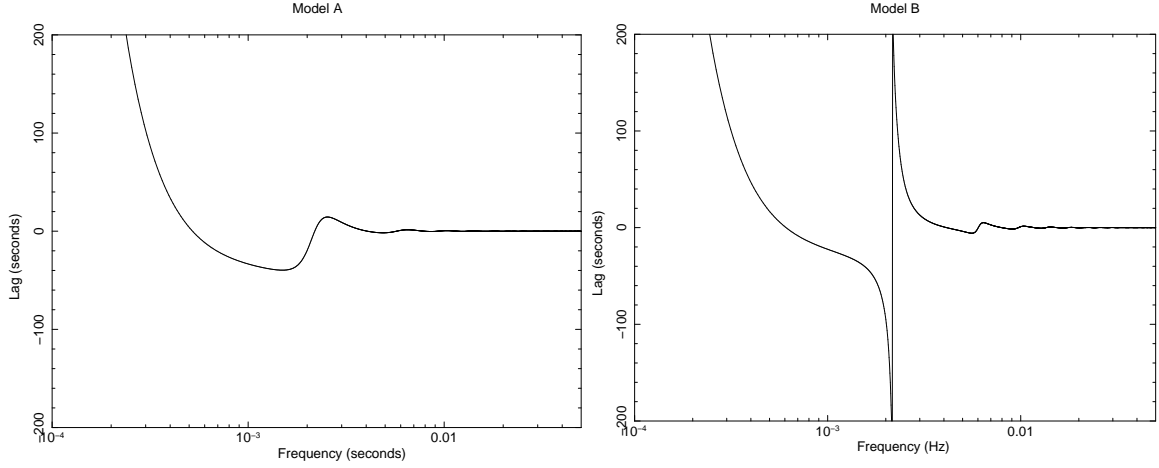

\includegraphics[bb=36 50 570 700,angle=270,width=0.5\textwidth]{MOD.FINAL3.model_kynrefrev.ps}
\includegraphics[bb=36 40 570 700,angle=270,width=0.5\textwidth]{MODHIGHSPIN.FINAL3.model_kynrefrev.ps}
\caption{Extrapolated to higher frequencies fitted models for 1H~0707-495 with the obtained value for spin given the data ($0.52{\pm}0.09$; model A) and for a highly spinning BH ($0.95$, model B) at left and right, respectively. }
\label{fig2b}
\end{figure}

We fitted the (0.3-1 versus 2-4\,keV) time-lag versus frequency global spectrum of 1H~0707-495 with the model {\tt KYNREFREV} and obtained a very good fit (${\chi}^{2}/{\nu}=28/30$; with ${\nu}$ being the number of d.o.f.)
as shown in Fig.~\ref{fig2}. The primary isotropic flux was set to slightly sub-Eddington (i.e. ${\rm L}/{\rm Ledd}=0.1$) according to previous studies \cite{fabian09}. The value for the density of the accretion
disc obtained was ${\rm N}_{\rm H}=(30{\pm}10){\times}10^{15}\,{\rm cm}^{-3}$ (adopting a constant radial density profile) and the values of the rest of the parameters of the fit (spin, mass and height of the lamp-post) are shown in 
Tab.~\ref{tab1}. The mass is in very good agreement with previous findings but the height of the lamp-post is higher ($5.8{\pm}0.5\,{\rm r}_{\rm g}$) than previously 
found ($2.4{\pm}0.5\,{\rm r}_{\rm g}$; \cite{emmanoulopoulos14}) and {\it the value for the BH spin is lower than usually obtained from X-ray spectroscopy} ($>0.98$, e.g. \cite{fabian09}) but in agreement with the previous value obtained from
time-lag fitting by \cite{emmanoulopoulos14} using a similar model. As shown in Fig.~\ref{fig2b} the difference between the obtained slow-rotation (model A) and high-spinning BH (model B) is that in the latter we obtain phase wrapping at 
high frequencies (still not detected with current detectors). The extrapolation of the model with high spin to the lower (observed) frequencies is not compatible with the data, hence the reason of getting a lower 
value for the spin of the BH.

\begin{figure}
\includegraphics[bb=0 -200 612 292,angle=270,width=0.4\textwidth]{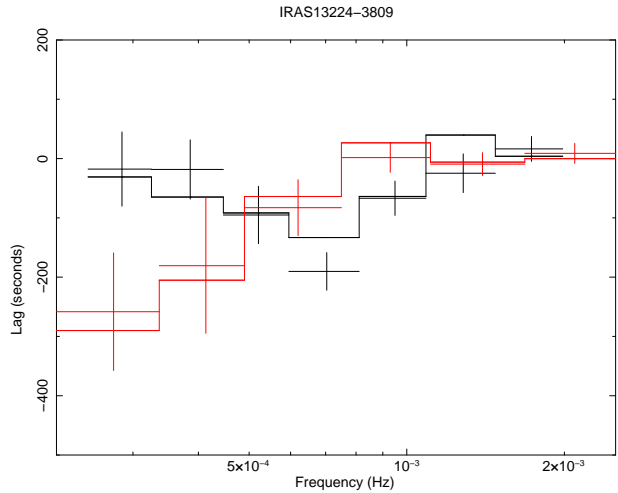}
\caption{The (0.3-1 versus 1.2-5\,keV) X-ray soft time-lag versus frequency spectrum of IRAS~13224-3809 fitted with the {\tt KYNREVREF} model in XSPEC (black and red line, for the low and the high-flux states, respectively).}
\label{fig3}
\end{figure}

When fitting the two datasets from IRAS~13224-3809 (the same as previously analyzed by \cite{kara13}) we obtained also a good fit (${\chi}^{2}/{\nu}=10/6$) as shown in Fig.~\ref{fig3}. We tested
the ``pure'' lamp-post geometry, in which the change of the state (i.e. luminosity) is due to the variable height of the corona. According to the expectation of a (near) Eddington accreting BH for the 
given mass \cite{boller97} we set the primary isotropic flux to a slightly sub-Eddington value (${\rm L}/{\rm Ledd}=0.1$). The value for the density of the accretion
disc obtained was ${\rm N}_{\rm H}=(26{\pm}5){\times}10^{15}\,{\rm cm}^{-3}$ (adopting a constant radial density profile) and the values of the rest of the parameters of the fit (spin, mass and heights of the lamp-post during
both states) are shown in Tab.~\ref{tab2}. We obtain a moderate value for the spin ($0.74{\pm}0.02$) that is lower than previously obtained ($a=0.99$ ; \cite{fabian12b})

\begin{figure}
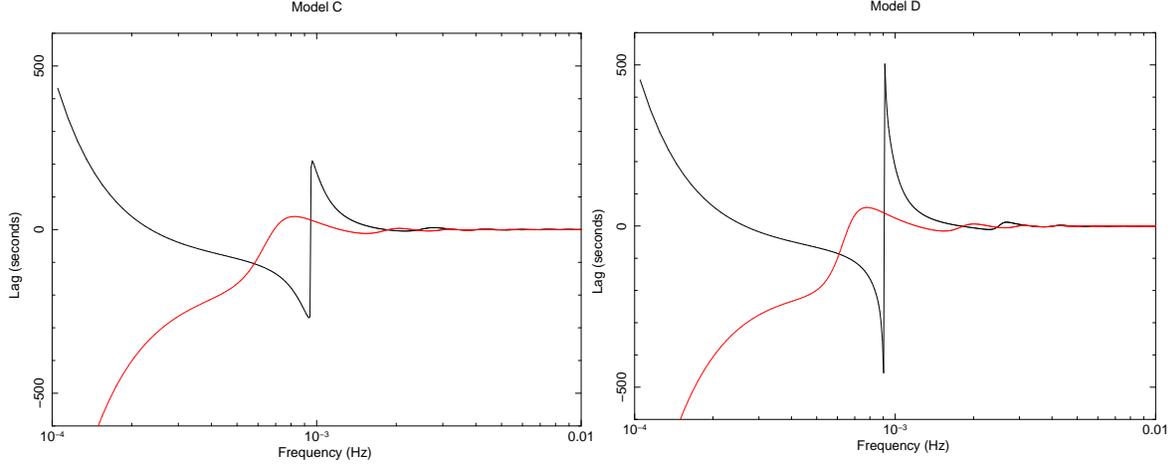

\includegraphics[bb=36 50 570 700,angle=270,width=0.5\textwidth]{MOD.BEST5.freeLFITTED.highlowfluxes.ps}
\includegraphics[bb=36 40 570 700,angle=270,width=0.5\textwidth]{MODHIGHSPIN.BEST5.freeLFITTED.highlowfluxes.ps}
\caption{Extrapolated to higher frequencies fitted models for IRAS~13224-3809 with the obtained value for spin given the data ($0.74{\pm}0.02$; model C) and for a highly spinning BH ($0.95$, model D) at left and right, respectively.}
\label{fig3b}
\end{figure}

The {\it reason for such a disagreement for the spin is the fact that
for such a high value (${\gtrsim}0.95$) phase-wrapping occurs (and it is more pronounced for high values of the intrinsic primary luminosity, after performing some testing)}. This can be seen in Fig.~\ref{fig3b} where the current fitted model (model C)
and a high-spinning BH (model D) are shown. If the spin is changed to a lower value (e.g. ${\le}0.5$) phase-wrapping totally disappears.

As pointed previously \cite{kara13} the change in the shape 
of the observed time-lags can be interpreted as a change in the properties of the corona. In the terms of the model and parametrization used this is shown in the form of a change in the height of the lamp-post 
(from $4.9{\pm}0.2\,{\rm r}_{\rm g}$ to $9.0{\pm}1.0\,{\rm r}_{\rm g}$). Both values are higher to what has been previously reported in the framework of the lamp-post model ($2.9{\pm}0.8\,{\rm r}_{\rm g}$; \cite{emmanoulopoulos14}), alike
in the case of 1H~0707-495, which might be due to the slightly different model used. Nevertheless, the values obtained for the Eddington ratio and coronal height are consistent with what is obtained and shown in Fig.~7 by 
recent work \cite{kara16}.

\begin{table}
\begin{tabular}{c c c c c c c}
\hline
 {\bf Mass} & {\bf Spin} & {\bf View. angle} & {\bf Height} & {\bf PL norm.}  &  {\bf PL index}  &  ${\chi}^{2}/{\rm d.o.f.}$  \\
  ${\rm M}$                                      &   ${\rm a}/{\rm M}$                         &  ${\theta}_{\rm o}$   &     $h$                                & ${\rm A(s)}$   &  ${\rm s}$   &            \\
             (${\times}10^6\,{\rm M}_{\odot}$)   &                      (${\rm GM}/{\rm c}$)   &                       &         (${\rm GM}/{\rm c}^{2}$)       &                &              &            \\
\hline
 $2.6{\pm}0.2$    & $0.52{\pm}0.09$       &  $77{\pm}3$    &  $5.8{\pm}0.5$    &  $(2.2{\pm}0.3){\times}10^{-6}$  &  $2.23{\pm}0.02$     &  28/30                            \\
\hline
\end{tabular}
\caption{Values of the main parameters obtained from the best-fit using the {\tt KYNREFREV} MODEL of the time-lags of 1H~0707-495 shown in Fig.~\protect\ref{fig2}. Errors are 68\,\% confidence errors. }
\label{tab1}
\end{table}

\begin{table}
\begin{tabular}{c c c c c c }
\hline
 {\bf Mass} & {\bf Spin} & {\bf View. angle} & {\bf Height (1)} & {\bf Height (2)}  &  ${\chi}^{2}/{\rm d.o.f.}$  \\
  ${\rm M}$\,(${\times}10^6\,{\rm M}_{\odot}$)   &   ${\rm a}/{\rm M}$\,(${\rm GM}/{\rm c}$)   &  ${\theta}_{\rm o}$   &     $h_1$ (${\rm GM}/{\rm c}^{2}$)       & $h_2$ (${\rm GM}/{\rm c}^{2}$)   &             \\
\hline
  $7.1{\pm}0.2$                  &   $0.74{\pm}0.02$     &    $66{\pm}5$          &   $4.9{\pm}0.2$    &  $9.0{\pm}1.0$                  &         10/6                       \\ 
\hline
\end{tabular}
\caption{Values of the main parameters obtained from the best-fit using the {\tt KYNREFREV} MODEL of the time-lags of IRAS~13224-3809 shown in Fig.~\protect\ref{fig3}. Errors are 68\,\% confidence errors. }
\label{tab2}
\end{table}

\section{Discussion and conclusions}  \label{discuss}

In this paper we briefly presented a model ({\tt KYNREFREV}) which deals with X-ray reverberation effects in AGN and can be used in and/or outside XSPEC. This model takes into account all the relativistic effects
acting on light in the vicinity of the BH and a realistic prescription of the accretion disc surrounding it under the lamp-post primary source geometry. A more detailed description of the model will be 
presented in a forthcoming publication \cite{dovciak17}. 

We present some preliminary interesting results obtained through the analysis of X-ray reverberation time-lags from 1H~0707-495 and IRAS~13224-3809 using this model. The global (0.3-1 versus ${\approx}2-4$\,keV) X-ray
soft time-lag spectrum is well fitted with this model, confirming previous statements on the X-ray reverberation origin of the negative X-ray soft time-lags \cite{fabian09}. We find that the BH is low and moderately-spinning 
for the case of 1H~0707-495 and IRAS~13224-3809, respectively. {\it This is 
in contrast with previous results obtained from X-ray spectroscopy which suggest a maximally-spinning BH for both sources}. We chose a near to Eddington value for the primary isotropic flux in the 2--10\,keV energy range of 
${\rm L}/{\rm L}_{\rm EDD}=0.1$, which is reasonable given the masses of both BHs. We notice that these results are sensitive to the chosen ${\rm L}/{\rm L}_{\rm EDD}$ value, mostly in the case of IRAS~13224-3809, with
a higher BH mass and hence with (observable with current instrumentation) phase-wrapping. We notice that, although our results are preliminary they are in agreement with previous claims of a low-spinning BH for the 
case of 1H~0707-495 \cite{done16}.

In the case of IRAS~13224-3809 we point out that the (0.3-1 vs 1.2-5\,keV) spectral time-lag transition related with the flux state of the source reported
previously \cite{kara13} might be due to a change in the height of the corona (from $4.9{\pm}0.2$ to $9.0{\pm}1.0\,{\rm r}_{\rm g}$ in the low and the high-flux states, respectively).
These results are consistent with the light-bending scenario proposed by \cite{miniutti04}, where in low-flux states the corona is closer to the BH and general relativistic effects are more important. If confirmed (with higher time-resolution
observations), these results would indicate the power of analyzing X-ray reverberation time-lags in order to set further constraints of the inner disc-corona geometry for these systems. 

%Further work is ongoing in order to prepare more detailed analysis of the obtained results. 

\acknowledgments

\noindent MCG and MD acknowledge support provided by the European Seventh Frame-work 
Programme (FP7/2007-2013) under grant agreement n$^{\circ}$ 312789. JS acknowledges financial support from 
the Grant Agency of the Czech Republic within the project n$^{\circ}$\,14-20970P.

\end{document}